\documentclass[twocolumn,showpacs,preprintnumbers,amsmath,amssymb,unsortedaddress,prb]{revtex4}

\usepackage{graphicx}
\usepackage{dcolumn}
\usepackage{bm}

\begin{document}


\title{Spin polarized transport through a single-molecule magnet:\\ current-induced magnetic switching}

\author{Maciej Misiorny}
 \email{misiorny@amu.edu.pl}
\author{J\'{o}zef Barna\'{s}}%
 \email{barnas@amu.edu.pl}
\affiliation{%
Department of Physics, Adam Mickiewicz University, 61-614
Pozna\'{n}, Poland
}%

\date{\today}

\begin{abstract}
Magnetic switching of a single-molecule magnet (SMM) due to
spin-polarized current is investigated theoretically. The charge
transfer between the electrodes takes place \emph{via} the lowest
unoccupied molecular orbital (LUMO) of the SMM. Generally, the
double occupancy of the LUMO level, and hence a finite on-site
Coulomb repulsion, is allowed. Owing to the exchange interaction
between electrons in the LUMO level and the SMM's spin, the latter
can be reversed. The perturbation approach (Fermi golden rule) is
applied to calculate current-voltage characteristics. The influence
of Coulomb interactions on the switching process is also analyzed.

\end{abstract}

\pacs{72.25.-b, 75.60.Jk, 75.50.Xx}


\maketitle

\section{Introduction}

Single-molecule magnets (SMMs) are molecules with a relatively large
net spin moment (corresponding to the spin number $S$) and a
significant uniaxial magnetic
anisotropy.~\cite{GatteschiAngewChem42/03,BlundellJPhysCondMatt16/04}
As a result, behavior of SMMs resembles much that of
superparamagnets, and at low temperatures the molecules become
trapped in one of the two metastable spin states $|\pm
S\rangle$.~\cite{SessoliNature365/93,GatteschiScience265/94,BarraEPL35/96,ThomasNature383/96}
Owing to this bistability, SMMs seem to be a suitable base for
memory cells of future information storage and processing
technology.~\cite{JoachimNature408/00,TimmPRB73/06} Apart from this,
SMMs can possibly become basic components of the molecular-based
spintronic devices.~\cite{WolfScience294/01}

Electronic transport through an individual SMM has been demonstrated
experimentally only very
recently,~\cite{HeerschePRL96/06,JoNanoLett6/06,HendersonCM/0703013}
attracting also a significant theoretical
attention.~\cite{KimPRL92/04,RomeikePRL96/06I,RomeikePRL96-97/06,ElstePRB73/06,MisiornyEPL78/07}
An important issue in this context is the question of how the
molecule's spin can be switched between the two stable states by
means of spin-polarized current. This question is important not only
from the purely fundamental reasons, but also from the point of view
of possible applications of SMMs in various magnetoelectronic
devices, and particularly as memory cells.

It is already well known that when a spin-polarized current flows
through a magnet, some amount of spin momentum can be transferred
from the current to magnetic body,~\cite{Ralph-Buhrman_book}
leading effectively to a spin-transfer torque. This additional
torque may lead to magnetic switching or current-induced
precessional states. It has been shown recently, that exchange
interaction between spin-polarized current and a SMM embedded in
the barrier of a magnetic tunnel junction can lead to reversal of
the molecule's spin.~\cite{MisiornyPRB75/07}  The model considered
there was simplified as the current was not flowing through the
molecule, but rather directly between magnetic electrodes.
However, the tunneling electrons could interact with the SMM {\it
via} the exchange coupling, leading to switching of the SMM. The
main objective of the present paper is to investigate
theoretically a more realistic mechanism of SMM's switching, when
the spin-polarized current flows directly through the molecule
(molecular single-electron transistor geometry). In the model
assumed, the current flows \emph{via} the lowest unoccupied
molecular orbital (LUMO) of the SMM. We restrict, however, our
consideration to  the case of the sequential transport regime. The
results clearly show that transport of electrons through the LUMO
level can lead to magnetic switching of the molecule, when the
electrons in the LUMO level interact \emph{via} exchange coupling
with the spin moment of the inner core of the SMM.

In Sec.~II we present the model and basis of the theoretical
analysis of transport characteristics. Numerical results are
presented in Sec.~III. These results clearly show the possibility
of magnetic switching induced by current pulse. Summary and final
conclusions are in Sec.~IV.

\section{Model and theoretical description}

\begin{figure}
\includegraphics[width=0.85\columnwidth]{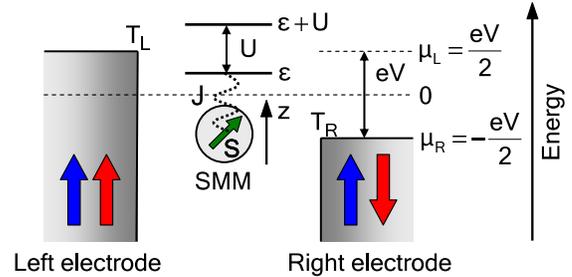}
\caption{\label{fig1}(Color online) Schematic representation of the
system under consideration in the nonequlibrium state, i.e. when a
finite bias voltage $V$ is applied, $eV=\mu_L-\mu_R$, where $\mu_L
(\mu_R)$ denotes the electrochemical potential of the left (right)
electrode. Two collinear magnetic configurations of the leads'
magnetic moments: parallel (black solid arrows) and antiparallel
(grey solid arrows) are also indicated.}
\end{figure}

The model to be considered, see Fig.~\ref{fig1}, consists of a SMM
weakly coupled to two ferromagnetic electrodes (also called here
leads). We assume that the electronic transport through the
molecule takes place only \emph{via} the LUMO level of energy
$\varepsilon$ (measured from the Fermi level of the leads at
equilibrium). This level is assumed to be exchange-coupled to the
SMM's spin.~\cite{TimmPRB73/06} Only collinear, i.e. parallel and
antiparallel configurations of the leads' magnetic moments are
considered, and these magnetic moments are assumed additionally to
be parallel to the magnetic easy-axis of the molecule (along the
axis \emph{z} in Fig.~\ref{fig1}).

The complete Hamiltonian of the system may be written as
$\mathcal{H}=\mathcal{H}_{S\! M\!
M}+\mathcal{H}_{el}+\mathcal{H}_T$. The first term describes the
SMM and is assumed in the form,
    \begin{align}
    \label{eq:HamSMM}
    \mathcal{H}_{S\! M\! M}& = -\Big(D +\! \! \sum_{\sigma=\{\uparrow,\downarrow\}}\! \! \Delta D_{1}\, c_\sigma^\dag
    c_\sigma^{} + \Delta D_{2}\, c_\uparrow^\dag c_\uparrow^{} c_\downarrow^\dag c_\downarrow^{}
    \Big)S_z^2
    \nonumber\\
    &+\! \! \sum_{\sigma=\{\uparrow,\downarrow\}}\! \! \varepsilon\,
    c_\sigma^\dag c_\sigma^{} + U\, c_\uparrow^\dag c_\uparrow^{} c_\downarrow^\dag c_\downarrow^{}
    \nonumber\\
    &-\frac{J}{2}\! \! \sum_{\sigma\sigma'=\{\uparrow,\downarrow\}}\! \! \bm{\sigma}_{\sigma\sigma'}\cdot\bm{S}\, c_\sigma^\dag
    c_{\sigma'}^{},
    \end{align}
where $\bm{S}$ is the SMM's spin operator, $\bm{\sigma}$ is the
Pauli spin operator for electrons in the LUMO level, and
$c_\sigma^\dag (c_\sigma^{})$ is the relevant electron creation
(anihilation) operator.  Apart from this, $U$ denotes the Coulomb
energy of two electrons of opposite spins in the LUMO level,
whereas $J$ is the exchange coupling parameter between the SMM's
spin and electrons in the LUMO level. The parameter $D$ is the
uniaxial anisotropy constant of a free molecule, while $\Delta
D_1$ and $\Delta D_2$ describe corrections to the anisotropy when
the LUMO level is occupied by one and two electrons,
respectively.~\cite{boukhvalov} The perpendicular anisotropy terms
have been omitted as irrelevant for the effects described here.
Apart from this, we neglect intrinsic spin relaxation, eg. that
due to spin-phonon coupling. The only spin relaxation taken into
account is that due to coupling of the molecule to the electrodes.
This is justified as spin relaxation due to electronic processes
associated with coupling of the dot to external leads is the
dominant one.

The next term of $\mathcal{H}$ describes the ferromagnetic
electrodes,
    \begin{equation}
    \label{eq:HamLeads}
    \mathcal{H}_{el} = \sum_{q=\{L,R\}}\sum_{{\bf k},
    \sigma=\{\downarrow,\uparrow\}}\varepsilon_{{\bf k}\sigma}^q\: a_{{\bf k}\sigma}^{q\dag}
    a_{{\bf k}\sigma}^q,
    \end{equation}
where $L (R)$ stands for the left (right) lead. The leads are
characterized by non-interacting electrons with the dispersion
relation $\varepsilon_{{\bf k}\sigma}^q$, where $\bf k$ denotes a
wave vector. In the equation above, $a^q_{{\bf k}\sigma}$ and
$a^{q\dag}_{{\bf k}\sigma}$ are the relevant annihilation and
creation operators, respectively.

The final term of the Hamiltonian $\mathcal{H}$ represents tunneling
processes between the leads and the molecule,
    \begin{equation}
    \label{eq:HamTun}
    \mathcal{H}_T = \sum_{q=\{L,R\}}\sum_{{\bf k},
    \sigma=\{\downarrow,\uparrow\}}\Big[T_q\, a_{{\bf k}\sigma}^{q\dag}c_\sigma^{} + T_q^* c_\sigma^\dag a_{{\bf
    k}\sigma}^q\Big],
    \end{equation}
where  $T_{L}$ and $T_{R}$ are the tunneling matrix elements
between the SMM and the left and right electrodes, respectively.
These parameters are assumed to be independent of the wave vector
and spin orientation. We point, that direct electron tunneling
between the leads is excluded.

It has been shown~\cite{TimmPRB73/06} that the Hamiltonian
$\mathcal{H}_{S\! M\! M}$, Eq.~(\ref{eq:HamSMM}), commutes with the
$z$ component $S_t^z$ of the total spin operator ${\bm S}_t \equiv
{\bm S} + \sum_{\sigma\sigma'}\bm{\sigma}_{\sigma\sigma'}\,
c_\sigma^\dag c_{\sigma'}^{}/2$, where the second term of $\bm{S}_t$
represents the spin of an electron in the LUMO. As a consequence, if
one treats $\mathcal{H}_{S\! M\! M}$ as the unperturbed part of the
total Hamiltonian $\mathcal{H}$, it is convenient to numerate the
eigenstates of $\mathcal{H}_{S\! M\! M}$ with the eigenvalues $m$ of
$S_t^z$ and with the number of electrons in the LUMO level. Thus,
the eigenstates of the SMM are given by:
$|0,m\rangle\equiv|0\rangle_o\otimes|m\rangle_{mol}$,
$|2,m\rangle\equiv|\!
\uparrow\downarrow\rangle_o\otimes|m\rangle_{mol}$,
$|1,m\rangle^\pm\equiv\mathbb{A}^\pm_m|\!
\downarrow\rangle_o\otimes|m+1/2\rangle_{mol}+ \mathbb{B}^\pm_m|\!
\uparrow\rangle_o\otimes|m-1/2\rangle_{mol}$ for the intermediate
states, and $|1,\pm S\pm 1/2\rangle\equiv|\!
\uparrow(\downarrow)\rangle_o\otimes |\pm S\rangle_{mol}$ for the
fully polarized states. According to our notation,
$|\bullet\rangle_{o(mol)}$ denotes the spin state of the orbital
(SMM). The coefficients $\mathbb{A}^\pm_m$ and $\mathbb{B}^\pm_m$
act here as effective Clebsch-Gordan coefficients which depend on
the system's parameters, and have the form
    \begin{align}
    \label{eq:EffClebschGordan}
    \mathbb{A}^\pm_m& = \mp \frac{\sqrt{2\Delta\epsilon(m)\pm
    (2D^{(1)}-J)m}}{2\sqrt{\Delta\epsilon(m)}},
    \\
    \mathbb{B}^\pm_m& =
    \frac{J\sqrt{S(S+1)-m^2+1/4}}{2\sqrt{\Delta\epsilon(m)}\sqrt{2\Delta\epsilon(m)\pm
    (2D^{(1)}-J)m}},
    \end{align}
where
$\Delta\epsilon(m)=\sqrt{D^{(1)}(D^{(1)}-J)m^2+(J/4)^2(2S+1)^2}$ and
$D^{(1)}\equiv D + \Delta D_1$. Additionally, we assume
$2D^{(1)}-J\geqslant 0$. The corresponding eigenenergies of the
Hamiltonian $\mathcal{H}_{S\! M\! M}$ are: $\epsilon(0,m)=-Dm^2$,
$\epsilon(2,m)=2\varepsilon+U-(D+2\Delta D_1+\Delta D_2)m^2$ and
$\epsilon(1,m)^\pm=\varepsilon+J/4-(D+\Delta
D_1)(m^2+1/4)\pm\Delta\epsilon(m)$. The energy of the fully
polarized states $|1,\pm S\pm 1/2\rangle$ is $\epsilon(1,\pm S\pm
1/2)^+$.

To investigate the current-induced magnetic switching of the SMM, we
analyze the relevant I-V characteristics. The total current flowing
through the molecule can be written as $I=(I_L-I_R)/2$, where
$I_\alpha$ $(\alpha=L,R)$ denotes the current flowing from the lead
$\alpha$ to the molecule,
    \begin{equation}
    \label{eq:TotalCurrent}
    I_\alpha=e\sum_{m_r,m_q}\sum_{n_r,n_q}
    (n_r-n_q)\gamma_\alpha^{|n_q,m_q\rangle|n_r,m_r\rangle}P_{|n_q,m_q\rangle}.
    \end{equation}
Here, $\gamma_\alpha^{|n_q,m_q\rangle|n_r,m_r\rangle}$ represents
the rate of transitions between the states $|n_q,m_q\rangle$ and
$|n_r,m_r\rangle$, whereas $P_{|n_q,m_q\rangle}$ is the
probability of finding the SMM in the state $|n_q,m_q\rangle$. We
assume that current is positive when electrons flow from the left
to right. For notational clarity, from now on we assume
$|n_q,m_q\rangle\equiv|q\rangle$, which also means that
$\sum_q\equiv\sum_{n_q}\sum_{m_q}$.

To find current we need to determine first both the transition
rates $\gamma_\alpha^{|q\rangle|r\rangle}$  and the probabilities
$P_{|q\rangle}$. Let us start with the transition rates. In the
second order (Fermi golden rule) one finds
$\gamma_\alpha^{|q\rangle|r\rangle}$ in the form,
    \begin{equation}
    \label{eq:TranRateFGR}
    \gamma_\alpha^{|q\rangle|r\rangle}=\sum_{k,\sigma\in\alpha}\Big[W^{k\sigma |q\rangle}_{|r\rangle}f(\varepsilon_{k\sigma}^\alpha)+
    W^{|q\rangle}_{k\sigma
    |r\rangle}\big[1-f(\varepsilon_{k\sigma}^\alpha)\big]\Big],
    \end{equation}
where the first term corresponds to electron transitions from the
$\alpha$-th lead to the molecule, while the second term describes
the charge transfer back to the lead $\alpha$. Furthermore,
$f(\varepsilon)$ is the Fermi-Dirac distribution function, and
$W_f^i=(2\pi/\hbar)|\langle
f|\mathcal{H}_T|i\rangle|^2\delta(E_f-E_i)$ is the rate of
transitions from an initial state ($i$) to a final state ($f$).

\begin{figure}
\includegraphics[width=0.8\columnwidth]{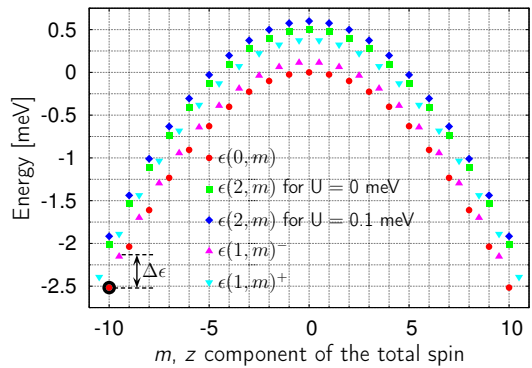}
\caption{\label{fig2} (Color online) The energy levels of a
$\textrm{Fe}_8$ molecular magnet for two values of the Coulomb
parameter $U$. The black bold circle indicates the initial state
$|0,-S\rangle$, and $\Delta\epsilon=0.36$ meV is the energy gap
corresponding to the activation energy for the magnetic switching. }
\end{figure}

The final expression for the transition rates
$\gamma_\alpha^{|q\rangle|r\rangle}$ takes the form
    \begin{align}
    \label{eq:TranRateFinal}
    \gamma_\alpha^{|q\rangle|r\rangle}&=\frac{1}{\hbar}\sum_{\sigma=\{\downarrow,\uparrow\}}\Gamma_\sigma^\alpha\Big\{
    \big|C_{qr}^\sigma\big|^2f\big(\epsilon(r)-\epsilon(q)-\mu_\alpha\big)\nonumber\\
    &\hspace{0cm} +
    \big|C_{rq}^\sigma\big|^2\Big[1-f\big(\epsilon(q)-\epsilon(r)-\mu_\alpha\big)\Big]\Big\},
    \end{align}
where $\Gamma_\sigma^\alpha=2\pi|T_\alpha|^2D_\sigma^\alpha$ is
the LUMO level width acquired due to coupling of the level to the
lead $\alpha$, and $D_\sigma^\alpha$ denotes the spin-dependent
density of states (DOS) at the Fermi level in the $\alpha$-th
electrode. These parameters will be used in the following to
describe strength of the coupling between the SMM and leads. It is
convenient to write $\Gamma_\sigma^\alpha$ as
$\Gamma^{\alpha}_{\pm}=\Gamma_{\alpha}(1\pm P_\alpha)$, where
$\Gamma_{\alpha}=(\Gamma^{\alpha}_{+}+\Gamma^{\alpha}_{-})/2$, and
$P_\alpha$ is the spin polarization of the lead $\alpha$,
$P_\alpha=(D_+^\alpha-D_-^\alpha)/(D_+^\alpha+D_-^\alpha)$. Here
$\sigma = +(-)$ corresponds to spin-majority (spin-minority)
electrons. In the following we assume that the couplings are
symmetric, $\Gamma_{\rm L}=\Gamma_{\rm R}= \Gamma/2$. Finally, in
Eq.~(\ref{eq:TranRateFinal}) $|C_{qr}^\sigma|^2\equiv|\langle
r|c^\dag_\sigma|q\rangle|^2$ together with
$|C_{rq}^\sigma|^2\equiv|\langle r|c_\sigma|q\rangle|^2$
constitute basic selection rules for transitions between
neighboring molecular states. The transition is allowed only when
the charge state of the SMM is changed by one and the change in
the total spin satisfies $\Delta S_t^z=\pm 1/2$. Assuming that the
SMM is  initially saturated in the state $|0,-S\rangle$, see
Fig.~\ref{fig2}, one may expect that at a sufficiently large
voltage the molecule can be switched to the final state
$|0,S\rangle$. The switching process corresponds then to the
reversal of the SMM's spin \emph{via} all the intermediate states.

The probabilities $P_{|q\rangle}$ (see Eq.~(\ref{eq:TotalCurrent})),
are obtained from the master equations
    \begin{equation}
    \label{eq:MasterEquation}
    c\, \frac{dP_{|q\rangle}}{dV}=\sum_{\alpha}\sum_r\Big[\gamma_\alpha^{|r\rangle|q\rangle}P_{|r\rangle}-
    \gamma_\alpha^{|q\rangle|r\rangle}P_{|q\rangle}\Big],
    \end{equation}
for $n_q=0,1^{(\pm)},2$ and $m_q\in \langle-S-1/2,S+1/2\rangle$
(we recall the definition, $|n_q,m_q\rangle\equiv|q\rangle$). In
the present paper, we assume that the voltage is augmented
linearly in time, $V=ct$, with $c$ denoting the speed at which the
voltage is increased. The corresponding time scale, however, is
much slower than that set by electronic transitions. The relevant
boundary conditions for the probabilities $P_{|q\rangle}$ are:
$P_{|0,-S\rangle}(V=0)=1$ and $P_{|q\rangle}(V=0)=0$ for
$|q\rangle\neq|1,-S\rangle$.

\section{Numerical results  and discussion}

The results have been computed for an octanuclear iron(III)
oxo-hydroxo cluster of the formula
$\left[\textrm{Fe}_8\textrm{O}_2(\textrm{OH})_{12}(\textrm{tacn})_6\right]^{8+}$
(shortly $\textrm{Fe}_8$). Such a free-standing molecule has spin
corresponding to $S=10$. The following values of the molecule
parameters are taken: $D=0.292$ K ($D\approx0.025$
meV),~\cite{WernsdorferScience284/99} $J=0.025$ meV and
$\varepsilon=0.25$ meV. Since there is no clear and reliable
experimental evidence of the change in anisotropy constant of the
$\textrm{Fe}_8$ molecule due to extra electrons in the LUMO level,
we assume $\Delta D_1 = \Delta D_2=0$. The couplings of the
molecule to the left and right lead are assumed to be the same,
$\Gamma_L=\Gamma_R=0.0015$ meV. We also assume that both the
electrodes are made of the metallic material characterized by the
same polarization parameter $P$, $P=P_L=P_R$. The calculations
have been performed for the temperature $T=0.01$ K, which is below
the blocking temperature $T_B=0.36$ K. The corresponding energy
levels of the molecule are shown in Fig.2. It is worth noting that
for the parameters assumed, the ground spin state of the molecule
attached to the leads ($S_t^z=\pm10$) is the same as that of a
free-standing molecule ($S_z=\pm10$). Nevertheless, for a
sufficiently low energy of the LUMO level, which can be controlled
for instance with a gate voltage, the ground state of the molecule
attached to the leads can correspond to $S_t^z=\pm 21/2$ (the
molecule with one extra electron on the LUMO level).

\begin{figure}
\includegraphics[width=0.99\columnwidth]{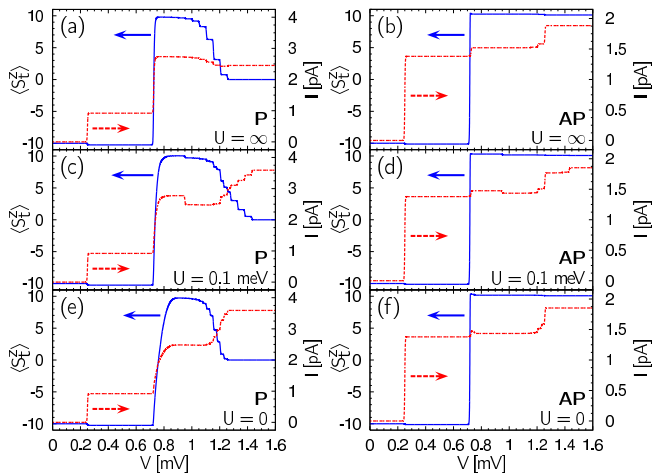}
\caption{\label{fig3} (Color online) The average value of the total
spin $\langle S_t^z \rangle$ (solid line) and the current $I$
flowing through the system (dashed line) in the case of parallel (P)
and antiparallel (AP) magnetic configurations for different values
of the Coulomb parameter $U$. The other parameters are:
$P_L=P_R=0.5$ and $c=1$ V/s.}
\end{figure}

Fig.~\ref{fig3} presents the average $\langle S_t^z\rangle$ and
current $I$ flowing through the system for different values of the
Coulomb parameter $U$ in both parallel and antiparallel magnetic
configurations of the leads. It can be noted that the reversal of
the SMM's spin occurs only in the antiparallel configuration,
whereas in the parallel configuration all molecular spin states
gradually become equally probable. As a consequence, $\langle
S_t^z\rangle\rightarrow0$ and the magnetic switching is not
observed. This is a consequence of the left/right symmetry of the
molecule's coupling to external leads -- similarly to the absence
of spin accumulation in tunneling through a metallic nanoparticle
in the parallel magnetic configuration. However, such a symmetry
is absent in the antiparallel configuration, and accordingly the
spin states of the molecule become unequally occupied, which in
turn results in spin reversal.

\begin{figure}
\includegraphics[width=0.95\columnwidth]{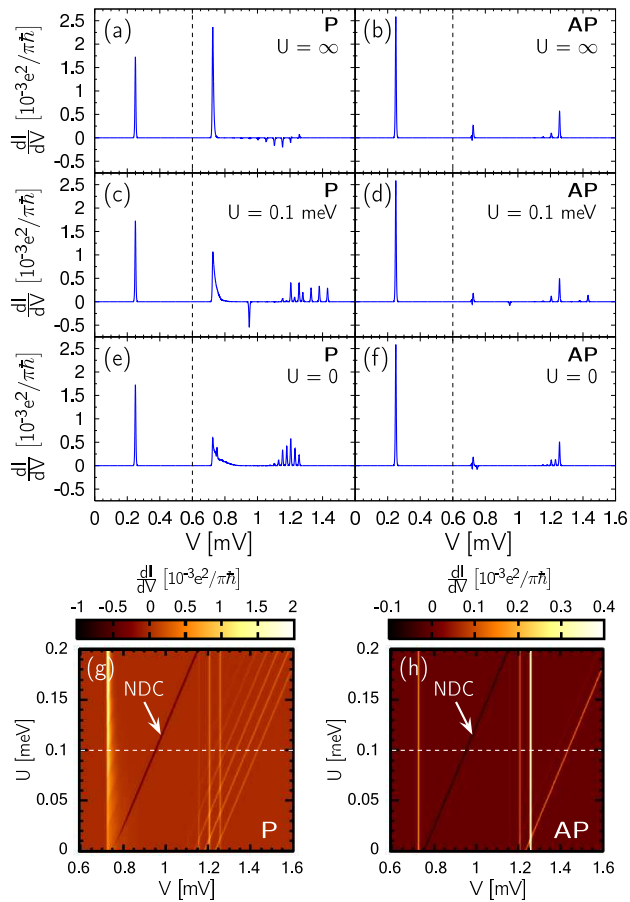}
\caption{\label{fig4} (Color online) Differential conductance
$dI/dV$ for two collinear, i.e. parallel (P) and antiparallel (AP)
magnetic configurations. The parameters as in  Fig.~\ref{fig3}.}
\end{figure}

The corresponding current-voltage characteristics, shown in
Fig.~\ref{fig3} by the dashed lines, reveal features (steps in the
current) which are directly related to the reversal process. In
fact, each step corresponds to a certain type of transitions
between neighboring molecular spin levels. Consider for instance
the main features of the $I$-$V$ plots shown in Fig.~\ref{fig3}.
The plots are almost the same as long as $V<0.72$ mV, i.e. for
$eV$ below the energy activating the reversal process. The first
step corresponds to the transition between the states
$|0,-10\rangle$ and $|1,-21/2\rangle$ (see also Fig.~\ref{fig2}).
As voltage increases further, the next step appears due to
transitions between the states $|0,-10\rangle$ and
$|1,-19/2\rangle^-$, and the magnetic switching begins. Augmenting
voltage further, one finds certain regions of bias voltage, where
current drops with increasing bias. This behavior is particularly
visible in the parallel configuration, see Figs.~\ref{fig3} (a,c).
The drop in current with increasing bias is equivalent to negative
differential conductance (NDC), see also Fig.~\ref{fig4}. The
negative differential conductance is a consequence of the spin
blockade phenomenon~\cite{WeinmannNATOASIserE291/95}, which in
turn follows from the inequality of the transition rates to the
two levels. The suppression of the current starts when the system
becomes energetically allowed to transfer from the state
$|1,-21/2\rangle$ to the state $|2,-10\rangle$.

The differential conductance corresponding to the $I$-$V$ curves
shown in Fig.~\ref{fig3} is presented in Fig.~\ref{fig4}. The
peaks correspond to the steps from Fig.~\ref{fig3}. The negative
differential conductance due to spin blockade is clearly seen,
particularly in the parallel configuration for $U>0$, although
some weak negative peaks also occur in the antiparallel case.
Evolution of the differential conductance with the Coulomb energy
$U$ and the bias voltage V is shown  explicitly in
Figs.~\ref{fig4} (g)-(h) for both magnetic configurations of the
leads.

\section{Summary}

In this paper we have considered electronic transport through a
single molecular magnet attached to ferromagnetic leads. The
molecule is characterized by a spin number $S$ and an additional
unoccupied orbital, which becomes active in transport through the
molecule.

We have shown that spin polarized electrons tunneling through the
LUMO level of a SMM can revers the SMM's spin when the electrons
in the LUMO level interact {\it via} exchange coupling with the
SMM's spin. The reversal starts at a certain threshold voltage
corresponding to the distance between the two lowest energy
levels. It is interesting to note, that for symmetrical systems,
the spin reversal takes place only in the antiparallel
configuration. The conductance spectra also show regions of
negative differential conductance due to spin blockade effect.

\begin{acknowledgements}

This work, as part of the European Science Foundation EUROCORES
Programme SPINTRA, was supported by funds from the Ministry of
Science and Higher Education as a research project in years
2006-2009 and the EC Sixth Framework Programme, under Contract N.
ERAS-CT-2003-980409.

\end{acknowledgements}


\begin{thebibliography}
\expandafter\ifx\csname
natexlab\endcsname\relax\def\natexlab#1{#1}\fi
\expandafter\ifx\csname bibnamefont\endcsname\relax
  \def\bibnamefont#1{#1}\fi
\expandafter\ifx\csname bibfnamefont\endcsname\relax
  \def\bibfnamefont#1{#1}\fi
\expandafter\ifx\csname citenamefont\endcsname\relax
  \def\citenamefont#1{#1}\fi
\expandafter\ifx\csname url\endcsname\relax
  \def\url#1{\texttt{#1}}\fi
\expandafter\ifx\csname
urlprefix\endcsname\relax\def\urlprefix{URL }\fi
\providecommand{\bibinfo}[2]{#2}
\providecommand{\eprint}[2][]{\url{#2}}

\bibitem{GatteschiAngewChem42/03}
D. Gatteschi  and R. Sessoli, Angew. Chem. Int. Ed. {\bf 42}, 268
(2003).

\bibitem{BlundellJPhysCondMatt16/04}
S.J. Blundell  and F.L. Pratt, J. Phys.: Condens. Matter {\bf 16},
R771 (2004).

\bibitem{SessoliNature365/93}
R. Sessoli, D. Gatteschi, A. Caneschi and M.A. Novak, Nature {\bf
365}, 141 (1993).

\bibitem{GatteschiScience265/94}
D. Gatteschi, A. Caneschi, L. Pardi and R. Sessoli, Science {\bf
265}, 1054 (1994).

\bibitem{BarraEPL35/96}
A.-L. Barra, P. Debrunner, D. Gatteschi, Ch.E. Schultz and R.
Sessoli, Europhys. Lett. {\bf 35}, 133 (1996).

\bibitem{ThomasNature383/96}
L. Thomas, F. Lionti, R. Ballou, D. Gatteschi, R. Sessoli and B.
Barbara, Nature {\bf 383}, 145 (1996).

\bibitem{JoachimNature408/00}
C. Joachim, J.K.  Gimzewski and A. Aviram, Nature {\bf 408}, 541
(2000).

\bibitem{TimmPRB73/06}
C. Timm and F. Elste, Phys. Rev. B {\bf 73}, 235304 (2006).

\bibitem{WolfScience294/01}
S.A. Wolf, D.D. Awschalom, R.A. Buhrman, J.M. Daughton, S. von
Moln\'{a}r, M.L. Roukes, A.Y. Chtchelkanova and D.M. Treger,
Science {\bf 294}, 1488 (2001).

\bibitem{RochaNatureMat4/05}
A.R. Rocha, V.M. Garcia-Su\'{a}rez, S.W. Bailey, C.J. Lambert, J.
Ferrer and S. Sanvito, Nature Materials {\bf 4}, 335 (2005).

\bibitem{HeerschePRL96/06}
H.B. Heersche, Z. de Groot, J.A. Folk, H.S.J. van der Zant, C.
Romeike, M.R. Wegewijs, L. Zobbi, D. Barreca, E. Tondello and A.
Cornia, Phys. Rev. Lett. {\bf 96}, 206801 (2006).

\bibitem{JoNanoLett6/06}
M.-H.  Jo, J.E. Grose, K. Baheti, M.M. Deshmukh, J.J. Sokol, E.M.
Rumberger, D.N. Hendrickson, J.R. Long, H. Park and D.C. Ralph,
Nano Lett. {\bf 6}, 2014 (2006).

\bibitem{HendersonCM/0703013}
J.J. Henderson, C.M. Ramsey, E. del Barco, A. Mishra and G.
Christou, J. Appl. Phys. {\bf 101}, 09E102 (2007).

\bibitem{KimPRL92/04}
G.-H. Kim  and  T.-S. Kim, Phys. Rev. Lett. {\bf 92}, 137203 (2004).

\bibitem{RomeikePRL96/06I}
C. Romeike,  M.R. Wegewijs and H. Schoeller, Phys. Rev. Lett. {\bf
96}, 196805 (2006).

\bibitem{RomeikePRL96-97/06}
C. Romeike, M.R. Wegewijs, W. Hofstetter and H. Schoeller, Phys.
Rev. Lett. {\bf 96}, 196601 (2006); {\bf 97}, 206601 (2006).

\bibitem{ElstePRB73/06}
F. Elste and C. Timm, Phys. Rev. B {\bf 73}, 235305 (2006); {\bf
75}, 195341 (2007).

\bibitem{MisiornyEPL78/07}
M. Misiorny and J. Barna\'{s}, Europhys. Lett. {\bf 78}, 27003
(2007).

\bibitem{Ralph-Buhrman_book}
D.C. Ralph and R.A. Buhrman, in {\it Concepts in spin
electronics}, edited by S. Maekawa (Oxford University Press,
2006), p. 195.

\bibitem{MisiornyPRB75/07}
M. Misiorny and J. Barna\'{s}, Phys. Rev. B {\bf 75}, 134425 (2007).

\bibitem{boukhvalov} K. Park and M.R. Pederson,  Phys. Rev. B {\bf 70}, 54414 (2004);
 D.W. Boukhvalov, M. Al-Sager, E.Z. Kurmaev, A. Moewes, V.R. Galakhov,
L.D. Finkelstein, S. Chiuzbaian, M. Neumann, V.V. Dobrovitski,
M.I. Katsnelson, A.I. Lichtenstein, B.N. Harmon, K. Endo, J.M.
North, and N.S. Dalal, Phys. Rev. B {\bf 75}, 14419 (2007).

\bibitem{WernsdorferScience284/99}
W. Wernsdorfer
and R. Sessoli, Science {\bf 284}, 133 (1999).

\bibitem{WeinmannNATOASIserE291/95}
D. Weinmann, W. H\"{a}usler, K. Jauregui and B. Kramer, in
\emph{Quantum Dynamics of Submicron Structures}, edited by H.A.
Cardeira, B. Kramer and G. Sch\"{o}n, NATO ASI Series E {\bf 291},
297 (1995).

\end{thebibliography}
\end{document}